\documentclass{elsart3-1}

\usepackage{amssymb}

\usepackage[english,francais]{babel}
\usepackage[latin1]{inputenc}
 \usepackage{amsmath}
 \usepackage{amsfonts}

\newtheorem{e-proposition}[theorem]{Proposition}

\newtheorem{e-definition}[theorem]{Definition\rm}

\newtheorem{theoreme}{Th\'eor\`eme}[section]

\newtheorem{proposition}[theoreme]{Proposition}

\renewcommand{\theequation}{\arabic{equation}}
\def\og{\leavevmode\raise.3ex\hbox{$\scriptscriptstyle\langle\!\langle$~}}
\def\fg{\leavevmode\raise.3ex\hbox{~$\!\scriptscriptstyle\,\rangle\!\rangle$}}

\journal{the Acad\'emie des sciences}
\begin{document}
% Vous pouvez mettre dans la prochain ligne la rubrique choisie
% (si vous la connaissez) - meme deux, format : Rubrique1/Rubrique2
\centerline{}
\begin{frontmatter}

% Titre, auteurs et adresses

% utiliser la commande \thanksref dans \title, \author ou \address
%     pour les notes en bas de page ;

% utiliser la commande \ead pour l'adresse e-mail de chaque auteur
%    (après la commande \auteur) ;

% \title{Title\thanksref{label1}}
% \thanks[label1]{}
% \author{Name\thanksref{label2}}
% \ead{email address}
%
% \thanks[label2]{}
% \address{Address\thanksref{label3}}
% \thanks[label3]{}
\selectlanguage{francais}
\title{A Simple Lack-of-Fit Test for Regression Models}

% utiliser les étiquettes pour indiquer l'adresse de chaque auteur,
%     s'il y a plusieurs adresses

% \author[label1,label2]{}
% \address[label1]{}
% \address[label2]{}

\author[authorlabel1]{Aubin Jean-Baptiste},
\ead{jean-baptiste.aubin@utc.fr}
\author[authorlabel2]{Leoni-Aubin Samuela}
\ead{samuela.leoni@insa-lyon.fr}

\address[authorlabel1]{Univ. de Technologie de Compiègne, Rue Personne de Roberval - BP 20529,
  60205 Compiègne, France.}
\address[authorlabel2]{INSA Lyon, ICJ, 20, Rue Albert Einstein,
 69621 Villeurbanne Cedex, France.}
% etc, etc

% Vous pouvez mettre a la prochaine ligne les dates
% (de reception et d'acceptation), et le nom du presentateur de votre Note

\medskip
\selectlanguage{francais}
\begin{center}
{\small Re\c{c}u le *****~; accept\'e apr\`es r\'evision le +++++\\
Pr\'esent\'e par £££££}
\end{center}

\begin{abstract}
% resume en francais, et apres l'abstract en anglais, qui
%    commence avec le titre en gras.
\selectlanguage{francais}
% Texte du résumé en français
\noindent {\bf Un test d'adéquation nonparamétrique pour la régression univariée. }\\

\noindent Dans le cadre de la régression univariée, nous proposons un outil nonparamétrique général permettant de tester si une fonction connue $m$ est un bon candidat pour la fonction de régression au vu des données. Ce test est basé sur la longueur maximale des suites ordonnées (par rapport à la covariable) des résidus de même signe. Aucune hypothèse n'est faite sur l'homoscédasticité des erreurs. De plus, ce test ne nécessite pas la présence de données répétées. Nous donnons ici la loi de la statistique test sous l'hypothèse nulle que la fonction considérée $m$ est la vraie fonction de régression ainsi que sous une certaine classe d'hypothèses alternatives.
{\it Pour citer cet article~: A. Nom1, A. Nom2, C. R.
Acad. Sci. Paris, Ser. I 340 (2005).}
\vskip 0.5\baselineskip

\selectlanguage{english}
\noindent{\bf Abstract}
\vskip 0.5\baselineskip
\noindent
A simple test is proposed for examining the correctness of a given completely specified response function against unspecified general alternatives in the context of univariate regression. The usual diagnostic tools based on residuals plots are useful but heuristic. We introduce a formal statistical test supplementing the graphical analysis. Technically, the test statistic is the maximum length of the sequences of ordered (with respect to the covariate) observations that are consecutively overestimated or underestimated by the candidate regression function.  Note that the testing procedure can cope with heteroscedastic errors and no replicates. Recursive formulae allowing to calculate the exact distribution of the test statistic under the null hypothesis and under a class of alternative hypotheses are given.%
{\it To cite this article: A. Nom1, A. Nom2, C. R.
Acad. Sci. Paris, Ser. I 340 (2005).}
\end{abstract}
\end{frontmatter}

% Maintenant la version abrégée en anglais, si présente
%\selectlanguage{francais}
%\section*{Abridged English version}

\selectlanguage{english}
% texte principale
\section{Introduction}
\label{}
\noindent Regression is one of the most widely used statistical tools to examine how one variable is related to another.
Statisticians usually begin their work by proposing a model for their observations. Then, they have to check on whether this model is correct. The graphical analysis of the residuals is an important step of this process since the detection of a systematic pattern would indicate a misspecified model. Unfortunately, this procedure is heuristic and could lead to errors of interpretation since it is often difficult to determine whether the observed pattern reflects model misspecification or random fluctuations. So it is of interest to complement such an analysis by a formal test. A large literature in this area can be found in Hart (1997). A review of statistical tests and procedures to determine lack of fit associated with the deterministic portion of a proposed linear regression model is  presented in Neill  and  Johnson (1984). We propose a new approach based on maximum length of sequences of consecutive overestimated (or underestimated) observations by the model. This test is very simple and can be computed visually if the sample size is small enough. This test is a modification of a nonrandomness test (see  Bradley 1968, chap. 11). In other words, we use this it to detect whether residuals are randomly distributed or not.\\
\noindent In Section 2,  the Length of the Longest Run Test is presented.  Section 3 is devoted to the law of the test statistic under the null hypothesis. In Section 4, the power of the test for a class of fixed alternatives is given.

\vskip 3mm

\section{The Length of the Longest Run Test Statistic}
%

 %      \subsection{Basic model}
\noindent Consider a collection of $n$ random variables $Y_i$ generated as
$$Y_i=m_0(x_i)+\varepsilon_i, ~~ i=1,\ldots,n, $$
\noindent where the $x_i$ are fixed design points and $m_0$ is the true regression function. Moreover, the $\varepsilon_i$ are independent and centered   random variables such that:
\begin{equation}
\label{condit}
\forall i=1,\ldots,n,~\quad Pr(\varepsilon_i>0)= Pr(\varepsilon_i<0)=\frac{1}{2}.
\end{equation}
\emph{\noindent Note that no hypothesis is made  on the regularity of the function $m_0$ or on the fact that errors must be identically distributed or homoscedastic, and that normality of $\varepsilon_i$ implies Condition (\ref{condit}). Moreover, contrary to other classical tests (like the F-test), no replicates are needed to compute our test statistic. 
}

% \medskip

\noindent We address the problem of testing the null hypothesis
$$H_0: m_0=m \qquad vs. \qquad H_1: m_0 \neq m ,$$
where $m$ is a completely specified function. \\
\noindent The $i$-th residual, $\widehat{\varepsilon}_i$, may be seen as substitute for the realisation of the random variable $\varepsilon_i$, thus comprising clues for adequacy or inadequacy of the model assumptions related to the distribution of $\varepsilon_i$. Some classical lack-of-fit test statistics are based on squared residuals, hence their signs are neglicted, and we can expect to loose some information. We propose a test statistic that takes these signs into account. This test statistic, $L_n$, is the maximum length of the sequences of ordered (with respect to the covariate) observations that are consecutively overestimated (or underestimated) by the candidate $m$. Formally, we define $Z_i:=\bf{1}_{\{\widehat{\varepsilon}_i>0\}}$, $1 \leq i\leq n$, $S_0:=0$, $S_l:=Z_1+\ldots+Z_l$, and put for $0 \leq K\leq n$,
\begin{equation*}
 I^+(n,K):= \displaystyle \max_{0 \leq l \leq n-K} (S_{l+K}-S_l).
\end{equation*}
\noindent Let $L_n^{+}$ be the largest integer $K$ for which $I^+(n,K)=K$. $L_n^{+}$ is the length of the longest run of 1's in $Z_1, \ldots, Z_n$, i.e. the length of the longest run of positive residuals. By analogy, we define $L_n^{-}$ as the length of the longest run of 0's in $Z_1, \ldots, Z_n$, that is $L_n^{-}$ is the largest integer $K$ for which $ I^-(n,K)=K$, where 
\begin{equation*}
 I^-(n,K):= \displaystyle \max_{0 \leq l \leq n-K} (K- S_{l+K}+S_l).
\end{equation*}
\noindent Clearly, $L_n^{-}$ is the length of the longest run of negative residuals. Finally, we define $L_n:=\max \left( L_n^{+},L_n^{-} \right).$

\noindent For a fixed nominal level $\alpha > 0$, we obtain the following unilateral rejection regions $W_{n,\alpha}=\left\{ L_n> c_{n,\alpha}\right\},$ where $c_{n,\alpha}$ is the largest integer such that $Pr(L_n> c_{n,\alpha}) \geq \alpha$. The corresponding bilateral rejection regions are $W_{n,\alpha}=\left\{ L_n \notin [c_{n,1-\alpha/2}, c_{n,\alpha/2}]\right\}$.

\vskip  3mm

\section{Distribution of $L_n$ under the null hypothesis}
\noindent If $m$  is equal to $m_0$, then, the residuals $\widehat{\varepsilon}_i$ \emph{are} the true errors $\varepsilon_i$. Since Condition (\ref{condit}) holds, we can apply  the following  recursive formula  (Riordan (1958),  p.153, Problem 13):
\begin{eqnarray*}
(n-1)!~Pr(L_n=k) & = & 2(n-2)!~Pr(L_{n-1}=k)-(n-k-2)!~Pr(L_{n-k-1}=k) \\% \label{rior}\\
& & +(n-2)!~Pr(L_{n-1}=k-1)-2(n-3)!~Pr(L_{n-2}=k-1) \\%\nonumber \\
& & +(n-k-1)!~Pr(L_{n-k}=k-1). %\nonumber
\end{eqnarray*}
\noindent By using $Pr(L_2=2)=1/2$ and $\forall ~ n > 0$, $Pr(L_n=1)= 1/2^{n-1}$, the entire exact distribution of $L_n$ and critical values for every nominal level can be deduced from the above formula.

\noindent  For most of practical cases of interest, $m$ is estimated. For example, if $m$ is estimated by OLS, an unfortunate property of residuals is that they are autocorrelated even when the true errors are white noise. This divergence from the assumptions disappears in large samples, but may be a problem when performing diagnostic tests in small samples. One way of handling this problem is to transform the OLS residuals so that they do satisfy the LS assumptions when these are correct. One of the most common of these transformations are the so called \emph{recursive residuals} (see Kianifard and Swallow (1996) among others).\\
Another possibility is to estimate $m$ on a subset of the data and to test it on the rest of the data.%\\

%  and this implies dependence of residuals. Nevertheless, if $m$ is consistently estimated, then asymptotical independence of residuals and Formula (\ref{rior}) still hold. 
%In other words, if $n$ is large enough, the probability distribution deduced from the above formula still gives a good %approximation for the true distribution.
\noindent In a coin tossing experiment, $L_n$, $L_n^+$, and $L_n^-$ can be seen as the length of the longest run of \emph{heads or tails},  \emph{heads} and  \emph{tails}, respectively. The length of the longest head run in a coin tossing experiment was investigated in the early days of probability theory. Later, Deheuvels (1985) gives upper and lower bounds for $L_n^+$ for a biased coin. 
\\ Schilling (1990) discusses the distributions  of $L_n$ for  unbiased coins, and remarks that for $n$ tosses of a fair coin the length of the longest run of \emph{heads or tails}, statistically speaking, tends to be about one longer than the length of the longest run of \emph{heads} only. For a biased coin, when $n$ is very large, if head is more likely than  tail, the distribution function of  $L_n^+$ is well approximated by an extreme value distribution (see Gordon et al. (1986)).

\section{Distribution of $L_n$ under fixed alternative hypotheses}
\noindent The distribution of the Length of the Longest Run Test statistic can be calculated under some fixed alternative hypotheses. First of all, we suppose that Condition (\ref{condit}) is fulfilled, and that errors are identically distributed.

\noindent Moreover, if we test 
$$ H_0: ~\forall~x,~m_0(x)=m(x) ~~~~~~vs.~~~~~~ H_{1,c}: ~\forall~x,~m_0(x)=m(x)+c,~~ c \neq 0$$
\noindent then, under $H_{1,c}$, the probability for an observation to be underestimated (respectively, overestimated), $p(c) \neq \frac{1}{2}$, is constant  
for all the observations. By considering the total number of positive residuals, $k$, in the sequence, the cumulative distribution  of $L_n$ can be expressed as :
$$ P(L_n \leq x) = \displaystyle \sum_{k=0}^n S_n^{(k)}(x) p(c)^k (1-p(c))^{n-k}, $$
\noindent where $S_n^{(k)}(x)$ is the number of sequences of length $n$ that contain $k$ positive residuals in which the length of the longest run of \textit{positive or negative residuals} does not exceed $x$. Analogously, Schilling (1990) studied the cumulative distribution of $L_n^+$.
 
\noindent In the following Proposition, we give a recursive formula to compute the $S_n^{(k)}(x)$:

\begin{proposition} 
\label{snk} Let $n$ and $x$ such that $0 < x \leq n$. Then, 

\begin{enumerate}
\item If $n-k \leq x$ and $ k \leq x$, 
% \begin{eqnarray}
$S_n^{(k)}(x)  = C^k_n$.
% \end{eqnarray}
\item If $n-k \leq x$ and $ k > x$, 
%\begin{eqnarray}
$S_n^{(k)}(x)  = \sum_{j=0}^x S_{n-j}^{(k)}(x)$.
%\end{eqnarray}
\item If $n-k > x$ and $ k \leq x$, 
%\begin{eqnarray}
$S_n^{(k)}(x)  = \sum_{j=0}^x S_{n-j}^{(k+1-j)}(x)$. 
%\end{eqnarray}
%
\item If $n-k > x$ and $ k > x$, 
let 
\begin{eqnarray}
R_n^{(k)}(x)  = \sum_{j\geq 0}  \left\{ \sum_{i=1}^x \right .& &  \left\{  S_{n-1-i-2j(x+1)}^{(k-1-j(x+1))}(x)+S_{n-1-i-2j(x+1)}^{(k-i-j(x+1))}(x) \right . \label{ricor} \\
& & \left. \left. - S_{n-1-(2j+1)(x+1)-i}^{(k-(j+1)(x+1))}(x) - S_{n-1-(2j+1)(x+1)-i}^{(k-1-j(x+1)-i)}(x) \right\}  \right\} \nonumber 
 \end{eqnarray}
\noindent with the following conventions: $\forall~x \in \mathbb{N}^*$, $R^{(0)}_0(x)=1$ and $\forall ~ n \in \mathbb{N}^*, ~ k \in \mathbb{N}^*, ~~R^{(-k)}_{-n}(x)=R^{(k)}_{-n}(x)=R^{(-k)}_{n}(x)=0$.
\noindent Finally, 
\begin{itemize}
\item If $ \exists ~ (i,j) \in \left\{1,\hdots,x \right\} \times \mathbb{N}^*$ such that 
$(k,n)=(2j(x+1)+i,j(x+1)) $ or $ (k,n)=(2j(x+1)+i,j(x+1)+i)$,
\noindent then $S_n^{(k)}(x) = R_n^{(k)}(x)+ 1$. 
\item If $ \exists ~ (i,j) \in \left\{1,\hdots,x \right\} \times \mathbb{N}^*$ such that 
$(k,n)=((2j+1)(x+1)+i,j(x+1)+i)$ or $ (k,n)=((2j+1)(x+1)+i,(j+1)(x+1))$,
\noindent then $S_n^{(k)}(x) = R_n^{(k)}(x)- 1$.
\item Else, $S_n^{(k)}(x) = R_n^{(k)}(x)$.
\end{itemize}
\end{enumerate}
\end{proposition}

\vspace{0.5cm}

\noindent  From this result, one can deduce the exact law of the test-statistic under $H_{1,c}$, and the power of the test follows. In the next Proposition, we show that, for $n$ large enough, the distribution function of $L_n$ is well approximated by the distribution function of $L_n^+$ (or $L_n^-$, depending on the value of $p(c)$):

\begin{proposition}
\label{muse}
If $\forall ~i=1, \ldots, n, ~ Pr(\varepsilon_i>0) = p(c), ~~ p(c)>\frac{1}{2},$ (resp. $p(c)<\frac{1}{2}$), then
$$\forall ~ k,~  Pr(L_n \leq k) = Pr(L_n^+ \leq k) + {o}(1) \textrm{ when } n \rightarrow \infty$$ 
\noindent (resp. $Pr(L_n \leq k) = Pr(L_n^- \leq k) + {o}(1)$).
\end{proposition}

%\vspace{0.5cm}

\section{Proofs.}

\noindent \textbf{Proof of Proposition \ref{snk}:}
\\ The recursive formula to compute $S_n^{(k)}(x)$, the number of sequences of length $n$ that contain $k$ positive residuals in which the length of the longest run of \textit{positive or negative residuals} does not exceed $x$, is found through a direct combinatorial analysis.
\\ We distinguish the following cases :\\

 \emph{(i)} For $n-k \leq x$ and $ k \leq x$, $ S_n^{(k)}(x)$ is equal to the binomial coefficient {\tiny $\left ( \begin{array}{c}
 n \\
 k \\
 \end{array}
 \right )$.}\\

 \emph{(ii)} When $n-k \leq x$ and $ k > x$, all the not-favorable sequences (that is, the sequences of length $n$ that contain $k$ positive residuals in which the length of the longest run of residuals having the same signs exceeds $x$) will contain at least a run of consecutively \emph{positive} residuals (and no run of consecutively \emph{negative} residuals) of length larger than $x$. In this particular case, we want to study the length of the longest head run in $n$ tosses of a biased coin including $k$ heads, problem solved by \cite{Schilling1990}. \\
 
 \emph{(iii)} In a similar way, when $n-k > x$ and $ k \leq x$, the problem is the same, swapping heads and tails.\\
 
 \emph{(iv)} 
For a fixed $x$ and $k$, when  $n-k > x$ and $ k > x$, The key is to partition the set of favorable sequences according to their beginning. Each sequence of length $n$ that contains $k$ positive residuals in which the length of the longest run of residuals having the same sign does not exceed $x$ can begin in at most $2x$ different ways 
and every beginning is followed by a sub-sequence with no more than $x$ consecutive residuals having the same sign. In  Table 1, %\ref{comi}
 we introduce the notation for the number of favorable sequences conditionally to the possible beginnings.

\begin{table}[ht]
\label{comi}
\centering
$\begin{array}{| c c c c c c c c l | c | c |}
\hline
1 & 2 & 3 & 4 & 5 & 6 & \ldots & x & (x+1)  & \textrm{Number of favorable sequences} &\textrm{Upper bound for the number of favorable sequences} \\
\hline
+ & - & 	&	&	&	&	&	&	& N_{1+-} &	 S_{n-2}^{(k-1)} (x)\\
+ & + & - & & & & & & &  N_{2+-} &S_{n-3}^{(k-2)} (x)\\
+ & + & + & -  & & & & & & N_{3+-} &  S_{n-4}^{(k-3)} (x) \\
+ & + & + & +  & -  & & & & &  N_{4+-} & S_{n-5}^{(k-4)} (x) \\
+ & + & + & +  & +  & - & & & &  N_{5+-} & S_{n-6}^{(k-5)} (x) \\
\vdots &  & 	&	&	&	&	&	&	&	\vdots & \vdots \\
+ & + & + & +  & +  & + & \cdots & + & - &  N_{x+-} & S_{n-(x+1)}^{(k-x)} (x)\\
- & + & 	&	&	&	&	&	&	&	 N_{1-+} & S_{n-2}^{(k-1)} (x) \\
- & - & + & & & & & & &  N_{2-+} & S_{n-3}^{(k-1)} (x) \\
- & - & - & +  & & & & & &  N_{3-+} & S_{n-4}^{(k-1)} (x) \\
- & - & - & -  & +  & & & & & N_{4-+} & S_{n-5}^{(k-1)} (x) \\
- & - & - & -  & -  & + & & & &  N_{5-+} & S_{n-6}^{(k-1)} (x) \\
\vdots &  & 	&	&	&	&	&	&	& \vdots &	\vdots \\
- & - & - & -  & -  & - & \cdots & - & + & N_{x-+} & S_{n-(x+1)}^{(k-1)} (x) \\
\hline
\end{array}$
\caption{The possible beginnings for a favorable sequence and the associated number of favorable sequencees (and upper bounds) 
}
\end{table}

\noindent Clearly, $S_{n}^{(k)} =N_{1+-} + N_{2+-} + \ldots + N_{x+-} + N_{1-+} + \ldots + N_{x-+}$. 
\\ Let determine the number of ``favorable'' sequences beginning by a positive residual and then a negative one $N_{1+-}$.\\

\noindent \emph{Step 1:}\\

\noindent $N_{+-}$ is at most equal to $S_{n-2}^{(k-1)} (x)$ (i.e., the number of favorable ways to complete a sequence beginning by $+-$, see Table 1). \\ %\ref{comi}).\\ 

\noindent \emph{Step 2:}\\

\noindent Among these $(n-2)-$sequences (the first two signs of the residuals are fixed), those beginning by $x$  ``-'' must be taken away because, in this case, the obtained sequences admit $x+1$ consecutive ``-'' (see Table 2).%\ref{tav1}).

\begin{table}[ht]\centering
$\begin{array}{ c c | c c c c | c }

1 & 2 & 3 & \cdots & (x+2) & (x+3) & \cdots\\
\hline
+ & - & - & \cdots & - & + & \cdots \\
\end{array}$
\label{tav1}
\caption{Form of the sequences to ``subtract'' to the $S_{n-2}^{(k-1)} (x)$ previous.}
\end{table}

\noindent There are $S_{n-2-(x+1)}^{(k-2)} (x)$ of them. At this point, $N_{+-}$ is at least equal to $ S_{n-2}^{(k-1)} (x) - S_{n-2-(x+1)}^{(k-2)} (x)$.\\

\noindent \emph{Further steps:}\\

\noindent Analogously, $(n-2-(x+1))-$sequences beginning by $x $ ``+'' must be subtracted from the $S_{n-2-(x+1)}^{(k-2)} (x)$ sequences taken away previously (see Table 3).

\begin{table}[ht]\centering
$\begin{array}{ c c | c c c c | c c c c | c}
1 & 2 & 3 & \cdots & (x+2) & (x+3) & (x+4) & \cdots & (2x+4) & (2x+5) & \cdots\\
\hline
+ & - & - & \cdots & - & + & + & \cdots & + & - & \cdots\\
\end{array}$
\label{tav2}
\caption{Form of the sequences to ``add'' to the $ S_{n-2}^{(k-1)} (x) - S_{n-2-(x+1)}^{(k-2)} (x)$ previous.}
\end{table}

\noindent Then $N_{+-} \leq  S_{n-2}^{(k-1)} (x) - \left( S_{n-2-(x+1)}^{(k-2)} (x) - S_{n-2-2(x+1)}^{(k-1-(x+1))} (x)\right) = S_{n-2}^{(k-1)} (x) - S_{n-2-(x+1)}^{(k-2)} (x) + S_{n-2-2(x+1)}^{(k-1-(x+1))} (x) $. Recursively,

$$ N_{+-} = \displaystyle \sum_{j \geq 0}  \left(  S_{n-2-(2j)(x+1)}^{(k-1-j(x+1))} (x) - S_{n-2-(2j+1)(x+1)}^{(k-2-j(x+1))} (x)  \right)$$

\noindent Note that for $j$ large enough, indexes become negative. We use the following conventions for all $x$, $S^{(0)}_0(x)=1$ and $\forall ~ n \in \mathbb{R}^*, ~ k \in \mathbb{N}^*, ~~S^{(-k)}_{-n}(x)=S^{(k)}_{-n}(x)=S^{(-k)}_{n}(x)=0$. We use the same method to calculate  $N_{..}$ for every possible beginning, we conclude the proof of Formula (\ref{ricor}) by summing them.
\\ There are some ``special points'' that need a correction when applying the Formula (\ref{ricor}). These points are such that the quantity $S_x^0(x)$ appears in the formula when $k < \frac{n}{2}$ (or the quantity $S_x^x(x)$ when $k > \frac{n}{2})$.
We underline that when $k=\frac{n}{2}$, the quantities $S_x^0(x)$ and $S_x^x(x)$ do not appear in the Formula.\\

\noindent For example, if $ \exists ~ (i,j) \in \left\{1,\hdots,x \right\} \times \mathbb{N}^*$ such that $(k,n)=(2j(x+1)+i,j(x+1))$, then the point $S_3^{(0)}(x)$ appears in the term $ - \sum_{i=1}^x S_{n-1-(2j+1)(x+1)-i}^{(k-(j+1)(x+1))}(x)$ of the recursive formula of $S_n^{(k)}(x)$. In this case, $S_x^{(0)}(x)$ represents the number of sequences of length $x$ with $x$ negative residuals and zero positive residuals that must be substracted when the last residual before the $x$ last ones is negative. So, this sequence ($S_x^{(0)}(x)=1$) mustn't be substracted since it hadn't been counted before (because it would have make appear a sequence of ($x$+1) consecutive negative residuals).\\

\noindent Note that in other cases, $S_x^{(0)}(x)$ has to be taken into account (for example is the $(n-x)$-sequence preceding the $x$ last residuals ends with a positive residual).\\

\noindent Similar considerations yield to the three other corrections.

\bigskip

\begin{table}[ht]\centering
$\begin{array}{| l | c c c c c c c c c c c c c c c c c c c |}
\hline
k & 0 & 1 & \cdots & \cdots & \cdots & \cdots & \cdots & \cdots & \cdots & \cdots & & (k-2-x) & (k-1-x) & (k-x) & \cdots & \cdots & (k-2) & (k-1) & (k) \\
\hline
n & & & & & & & & & & & & & & & & & & & \\
0 & & & & & & & & & & & & & & & & & & & \\
1 & & & & & & & & & & & & & & & & & & & \\
\vdots & & &\ddots &  & \vdots & & & & & & & & & & & & & & \\
\vdots & & & &+ &+ & & & & & & & & & & & & & & \\
\vdots & & & & &++ & & & & & & & & & & & & & & \\
\vdots & & & & & & & & & & & & & & & & & & & \\
\vdots & & & & & & & --& & & & & & & & & & & & \\
\vdots & & & & & & & - & - & & & & & & & & & & & \\
\vdots & & & & & & & \vdots & & &  \ddots & &  &  &  &  &  &  & &  \\
n-2- 3(x+1) & & & & & & & -      & & & & - &  &  &       &  &  &  & &  \\
  & & & & & & & & & & & & & & & & & & &\\
 n-2-3x-1   & & & & & & & + &  &   & & & + &  & & & &  & & \\
 \vdots &  & & & & & & & + &  & &  & + & &   & & & & & \\
  \vdots & & & & & & & & & & \ddots& & \vdots & & & & & & & \\
  & &  & & & & & & & & & +& + &  & &  & & & &\\
n-2-2x-2   & & & & & & & & & &  & & ++ & &  &  & & & & \\
n-2-2x-1   & & & & & & & & & & & & & &  &  & & & & \\
n-2-2x     & & & & & & & & & & & & & & - - &  & & & & \\
\vdots    & & & & & & & & & & & & & & -  & - & & & & \\
\vdots    & & & & & & & & & &  &  & & & \vdots &  & \ddots &  & &  \\
n-2-(x+1) & & & & & & & & & &  &  & & & -      &  &  & - & &  \\
 n-2-x & & & & & & & & & & & & & & & & & & &\\
 n-1-x & & & & & & & & & & & & & & + &  &  & & + & \\
 n-x & & & & & & & & & &  & & & & & + &  & & + & \\
  \vdots & & & & & & & & & & & & & & & & \ddots& & \vdots & \\
 n-3 & & & & & & & & & & & & & & & & & +& + & \\
 n-2 & & & & & & & & & & & & & & & & & & ++ & \\
 n-1 & & & & & & & & & & & & & & &  & & &  & \\
 n   & & & & & & & & & & & & & & & & &  & & \\
 \hline
\end{array}$
\label{tavola}
\caption{Illustration of Recursive Formula (\ref{ricor}), for a fixed $x$.}
\end{table}

\noindent The recursive formula (\ref{ricor}) becomes more clearful if we look at the Table (\ref{tavola}) which illustrates it.
\\ In Table (\ref{tavola}), for  fixed $n$, $k$ and $x$ we represent the coefficients to assign to each $S_{\bar{n}}^{(\bar{k})}(x)$ (where $\bar{k} < k$ and $\bar{n}<n$) in order to compute $S_n^{(k)}$. The sign ``$+$'' means that such coefficient equals 1, ``$-$'' that such coefficient equals $-1$, and an empty cell means that the coefficient equals 0.

\vspace{0.5cm}
\noindent \textbf{Proof of Proposition \ref{muse}:}
\\ \noindent This proposition is a direct application of the fact that $Pr(L_n^- < L_n^+)$ tends to $1$ when $n$ tends to infinity as shown in Muselli (2000), since, for all $x\geq 1$,
$$Pr(L_n \leq x ) = Pr(L_n <x | L_n^- < L_n^+) Pr(L_n^- < L_n^+) +Pr(L_n <x | L_n^- \geq L_n^+) Pr(L_n^- \geq L_n^+)$$ and 
$L_n = \max (L_n^-, L_n^+)$.

%\section*{Remerciements}
% Remerciements - texte ici


\begin{thebibliography}{5}
% Essayez à utiliser le systeme 'bibitem',
%    avec les references en ordre alphabetique.
% \bibitem{label1}
% Texte

%\bibitem{label2}
%
%\bibitem{label}
% \vskip 3mm
 


 
% \noindent 
 \bibitem{Bradley1968}
\textsc{Bradley, J. V.}
\textit{Distribution-free statistical tests}.
Prentice-Hall Inc,  1968. 


% \vskip 3mm
 

 \vskip 3mm
 
 \noindent 
 \bibitem{Deheuvels1985}
\textsc{Deheuvels, P.} 
On the Erdos-Renyi theorem for random fields and sequences and its relationships with the theory of runs and spacings.
\textit{Z. Wahrsch. Verw. Gebiete}   
\textbf{70} (1985) 91--115.  






 	
 	\vskip 3mm
 
  \noindent 
 \bibitem{Schillingal1986}
\textsc{Gordon, L., Schilling, M.F. and Waterman, M.S.}  
An Extreme value Theory for Long Head Runs
\textit{Probab. Th. Rel. Fields}   
\textbf{72} (1986) 279--287.  


 \vskip 3mm
 
  \noindent 
 \bibitem{Hart1997}
\textsc{Hart, J.}  
\textit{Nonparametric Smoothing and Lack-of-Fit Tests}.  
New York: Springer-
Verlag, 1997.

 \vskip 3mm


\noindent 
\bibitem{KianieSwall1996}
\textsc{Kianifard, F.}  and \textsc{Swallow, W. H.}
A review of the development and application of recursive residuals in linear models. 
\textit{J.A.S.A.}
\textbf{91} (433) (1996) 391--400.

\vskip 3mm 
 
 \noindent 
 \bibitem{Muselli2000}
\textsc{Muselli, M.}  
Useful inequalities for the longest run distribution.
\textit{Statistics \& Probability Letters}  
\textbf{46}   (2000)   239--249.  
	
 \vskip 3mm
 
 \noindent 
 \bibitem{Neillal1984}
\textsc{Neill, J. W.} and \textsc{Johnson, D. E.} 
Testing for lack of fit in regression -- A review.
\textit{Comm. Statist. A---Theory Methods}  
\textbf{13} (4)  (1984)   485--511.  
	
 \vskip 3mm
 

 
 \noindent 
 \bibitem{Riordan1958}                      % label for reference in the text
\textsc{Riordan, J.} 
\textit{An introduction to combinatorial analysis}.
John Wiley and sons, Inc, 1958.
 	
 \vskip 3mm
 
 \noindent 
 \bibitem{Schilling1990}
\textsc{Schilling, M. F.} 
 The longest run of heads. 
\textit{College Math. J.}  
\textbf{21} (1990) 196--207. 
 	

\end{thebibliography}
\end{document}